\documentclass[conference]{IEEEtran}
\IEEEoverridecommandlockouts

\usepackage{cite}
\usepackage{amsmath,amssymb,amsfonts}
\usepackage{algorithmic}
\usepackage{graphicx}
\usepackage{textcomp}
\usepackage{xcolor}
\def\BibTeX{{\rm B\kern-.05em{\sc i\kern-.025em b}\kern-.08em
    T\kern-.1667em\lower.7ex\hbox{E}\kern-.125emX}}

\usepackage{fancyhdr}
\usepackage{algorithmic}
\usepackage{graphicx}
\usepackage{textcomp}
\usepackage{xcolor}
\usepackage{graphicx}
\usepackage{subcaption}
\usepackage{tabularx}
\usepackage{float} 
\usepackage{multirow}
\usepackage{color}
\usepackage{arydshln} 
\usepackage{makecell}
\usepackage{arydshln} 
\usepackage{booktabs}

\usepackage[utf8]{inputenc}
\usepackage{tcolorbox}

\usepackage{booktabs} 

\makeatletter
\newcommand\thickhline{%
    \noalign{\ifnum0=`}\fi\hrule \@height 1pt \futurelet
    \reserved@a\@xhline}
\makeatother

\def\BibTeX{{\rm B\kern-.05em{\sc i\kern-.025em b}\kern-.08em
    T\kern-.1667em\lower.7ex\hbox{E}\kern-.125emX}}

\definecolor{Color1}{RGB}{110, 0, 110}   
\definecolor{Color2}{RGB}{110, 110, 0}    
\definecolor{Color3}{RGB}{0, 120, 140}    
\definecolor{Color4}{RGB}{150,150,128}

\begin{document}

\title{Knowledge-based Emotion Recognition using \\ Large Language Models}

\author{\IEEEauthorblockN{Bin Han}
\IEEEauthorblockA{\textit{Institute for Creative Technologies} \\
\textit{University of Southern California}\\
Los Angeles, CA, USA \\
binhan@usc.edu}
\and
\IEEEauthorblockN{Cleo Yau}
\IEEEauthorblockA{\textit{California State Polytechnic} \\
\textit{University-Pomona}\\
Pomona, CA, USA \\
cleoyau2020@gmail.com}
\and 
\IEEEauthorblockN{Su Lei}
\IEEEauthorblockA{\textit{audEERING GmbH} \\
Gilching, Germany \\
slei@audeering.com}

\and
\IEEEauthorblockN{Jonathan Gratch}
\IEEEauthorblockA{\textit{Institute for Creative Technologies} \\
\textit{University of Southern California}\\
Los Angeles, CA, USA \\
gratch@ict.usc.edu}
}

\maketitle
\thispagestyle{fancy}

\begin{abstract}
Emotion recognition in social situations is a complex task that requires integrating information from both facial expressions and the situational context. 
While traditional approaches to automatic emotion recognition have focused on decontextualized signals, recent research emphasizes the importance of context in shaping emotion perceptions. 
This paper contributes to the emerging field of context-based emotion recognition by leveraging psychological theories of human emotion perception to inform the design of automated methods. 
We propose an approach that combines emotion recognition methods with Bayesian Cue Integration (BCI) to integrate emotion inferences from decontextualized facial expressions and contextual knowledge inferred via Large-language Models. 
We test this approach in the context of interpreting facial expressions during a social task, the prisoner's dilemma. 
Our results provide clear support for BCI across a range of automatic emotion recognition methods. The best automated method achieved results comparable to human observers, suggesting the potential for this approach to advance the field of affective computing.
\end{abstract}


\section{Introduction} 
People readily make inferences about others from emotions expressed in social situations and use these inferences to guide social actions. Yet the field of affective computing has struggled to endow machines with this basic level of emotional intelligence. 
Recent research has highlighted the importance of situational knowledge in shaping emotion perceptions~\cite{kosti2020context}. 
Whereas automatic emotion recognition has traditionally focused on recognizing emotions from decontextualized signals (e.g., facial expressions labeled without knowledge of the situation that evoked the expression~\cite{bargal2016emotion, canal2022survey}), research on human social cognition highlights that people integrate cues from both expressions and their rich understanding of the social situation~\cite{carroll1996facial}. Yet “context-based” emotion recognition is still in its infancy. This paper contributes to this growing field by demonstrating how psychological theories of human emotion perception can inform the design of automated methods.

There are obvious benefits if emotion perceptions could be predicted from decontextualized signals alone, as was claimed possible by early psychological theories
\cite{ekman1971constants,ekman1992argument}.
Expressions could be easily collected and annotated without regard for context and the resulting algorithms utilized in any domain. 
Unfortunately, it is now clear that impressions formed from de-contextualized expressions often have little bearing on predicting what people feel, nor can they predict the inferences of observers knowledgeable of the social situation~\cite{barrett2019emotional}.

As a consequence, knowledge of the situational context must be incorporated into the recognition process. 
One obvious approach is to train recognition methods for specific contexts (e.g., emotional states of a driver~\cite{naqvi2020deep} or patient in a mental health screening~\cite{turabzadeh2018facial}), but this limits the generality of the resulting algorithm to these specific contexts. As an alternative, recent computer vision approaches have tried to infer the context by examining information in the background of an image or video (e.g., recognizing that a particular expression was produced in the context of a birthday party~\cite{kosti2017emotion}). Unfortunately, the information that can be inferred in this way is often quite limited. In contrast, people engaged in social interactions often have rich semantic knowledge about the nature of the shared task including each party's recent actions. 

Two recent innovations suggest how to incorporate situational knowledge while maintaining the advantages of decontextualized emotion recognition. 
First, research indicates people infer emotion from expressions using context-free methods and adjust based on situational knowledge. 
For example, Ong and colleagues’ Bayesian Cue Integration (BCI) model shows that context-specific emotion judgments can be decomposed into judgments based on the expression alone and from the situation alone, then integrated via Bayesian inference~\cite{ong2015affective} (see~\cite{houlihan2023emotion} for a similar approach). 
This implies that affective computing could similarly decompose the problem: i.e., utilize existing context-free emotion recognition methods to recognize emotions from expressions alone and then “post-process” the output of these algorithms assuming it was possible to make emotional inferences about situations. 

\begin{figure*}
    \centering
    \includegraphics[width=0.90\linewidth]{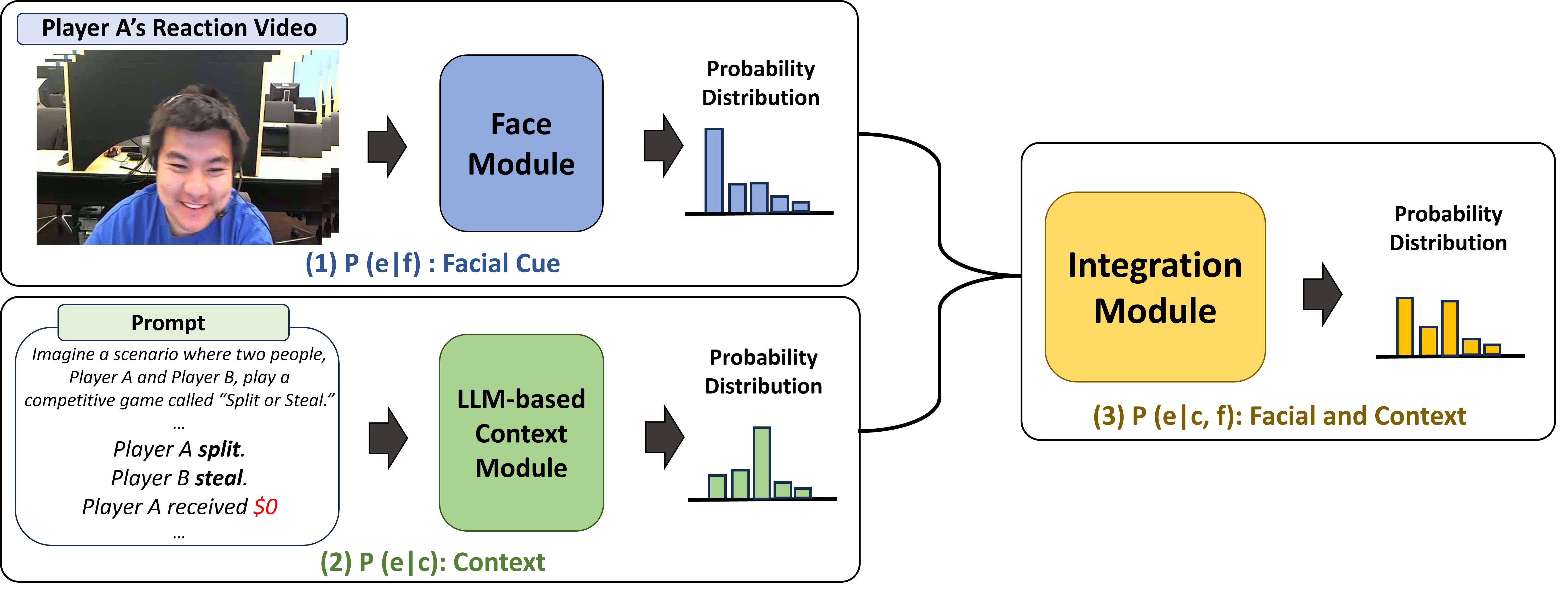}
    \caption{Illustrates knowledge-based recognition: (1) an emotion distribution is estimated from facial cues alone (2) from knowledge of the situational context (3) then integrated to predict context-based emotion perceptions.}
    \label{fig:overview}
\end{figure*}

Second, research into the emotional reasoning capabilities of Large Language Models (LLMs) suggests they are surprisingly accurate at predicting the emotions people feel across a wide range of situations. 
For example, Tak and Gratch found that GPT models accurately predict human emotional responses and appraisals across a wide range of situations~\cite{tak2023gpt}, and Broekens demonstrated GPT's zero-shot abilities in tasks such as sentiment analysis and appraisal-based emotion elicitation~\cite{broekens2023fine}. Additionally,  Resendiz and Klinger's work on emotion-conditioned text generation further highlights LLMs' capabilities in handling emotion-related tasks~\cite{resendiz2023emotion}.

Together, this suggests a general approach to context-dependent emotion recognition: First, predict the emotions people are likely to perceive from an emotional expression without context. Second, predict the emotions people are likely to perceive from a situational description. Finally, combine these separate sources of information with psychologically-inspired models such as BCI (see Fig.~\ref{fig:overview}).

We test this idea by examining how observers interpret facial expressions produced during an emotional social task (playing the prisoner’s dilemma game for money). 
We first replicate prior findings that human observers need context (emotion ratings from decontextualized videos differ considerably from ratings when fine-grained details of the context are provided). 
We next systematically explore the utility of BCI for fully-automated methods. Specifically, we apply the Bayesian approach to several context-free emotion recognition algorithms to assess the generality of the approach. We further investigate alternative LLMs for their ability to reason about emotional situations. Finally, we contrast BCI with alternative integration methods.  
To preview, our results provide clear support for BCI. The method improved accuracy across all of the context-free methods we tested. Second, GPT-4 was found to be an effective method for predicting emotions from situations. The best-performing results achieved human-level performance (as judged by comparing it with the predictions of BCI using human, rather than machine judgments),  We discuss these results and future directions, including the need to verify these findings on a broader range of situations.

\section{Bayesian Cue Integration}
\label{sec:bayesian_cue_integration}

We first introduce BCI~\cite{ong2015affective} and illustrate how it captures human judgments in the Prisoner's Dilemma task before turning to automated approaches. BCI predicts \textit{context-based} emotion judgments (i.e., judgments by human observers with extensive knowledge of the social context) from \textit{context-free} judgments (i.e., judgments by human observers without any knowledge of the context) and \textit{context-only} judgments (i.e., human judgments based on information about the context but without knowledge of the expression. Each of these judgments is represented as a probability distribution (i.e., the probability that a given human observer would make this judgment). The model assumes observers employ an intuitive theory for interpreting expressive and contextual cues: observers assume that the outcome of a social task (e.g., the joint decision in the prisoner's dilemma), influences emotions, which in turn affects facial expressions. The equation below captures this assumption (see~\cite{ong2015affective} for details on its derivation),


\begin{equation}
\resizebox{.5\hsize}{!}{$P(e|c,f) \propto \frac{P(e|f)P(e|c)}{P(e)}$}
\label{eq:bayesian}
\end{equation}


In Eq (1), $P(e|f)$ is the probability of reporting someone feels an emotion from the face alone. 
$P(e|c)$ is the probability that someone would be rated as experiencing an emotion given only the situation (e.g., without seeing a person’s face, how likely is a person to experience joy if we know they were just exploited). 
$P(e)$ is the a priori probability that different emotions tend to occur. 

\begin{figure}
    \centering
    \includegraphics[width=0.90\linewidth]{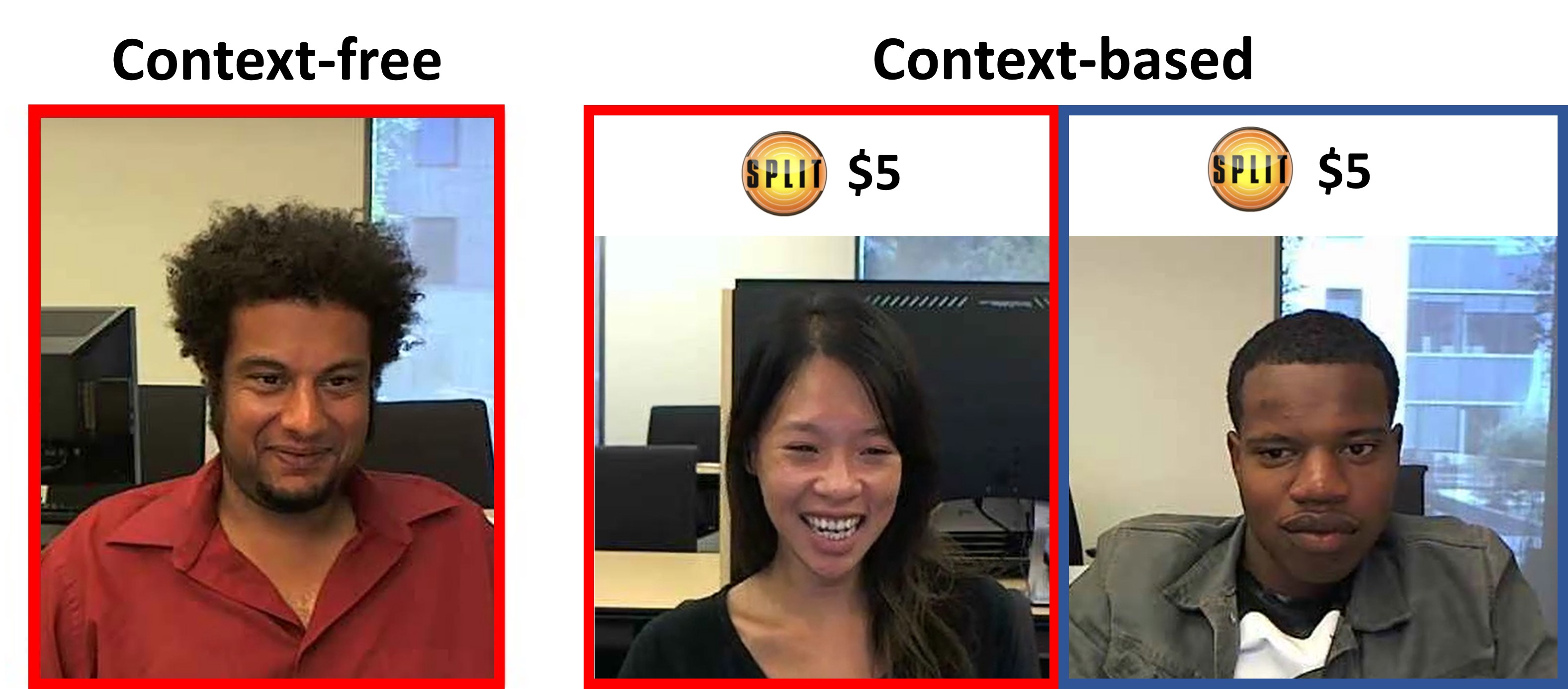}
    \caption{Facial reactions were annotated either without context (\textit{context-free}), with instruction about the joint outcome of the game (in this case mutual cooperation) seeing the reaction of both players reactions (\textit{context-based}).}
    \label{fig:context_level}
\end{figure}

\begin{figure*}[h]
    \centering
    \begin{subfigure}[b]{0.389\textwidth}
        \centering
        \includegraphics[width=\textwidth]{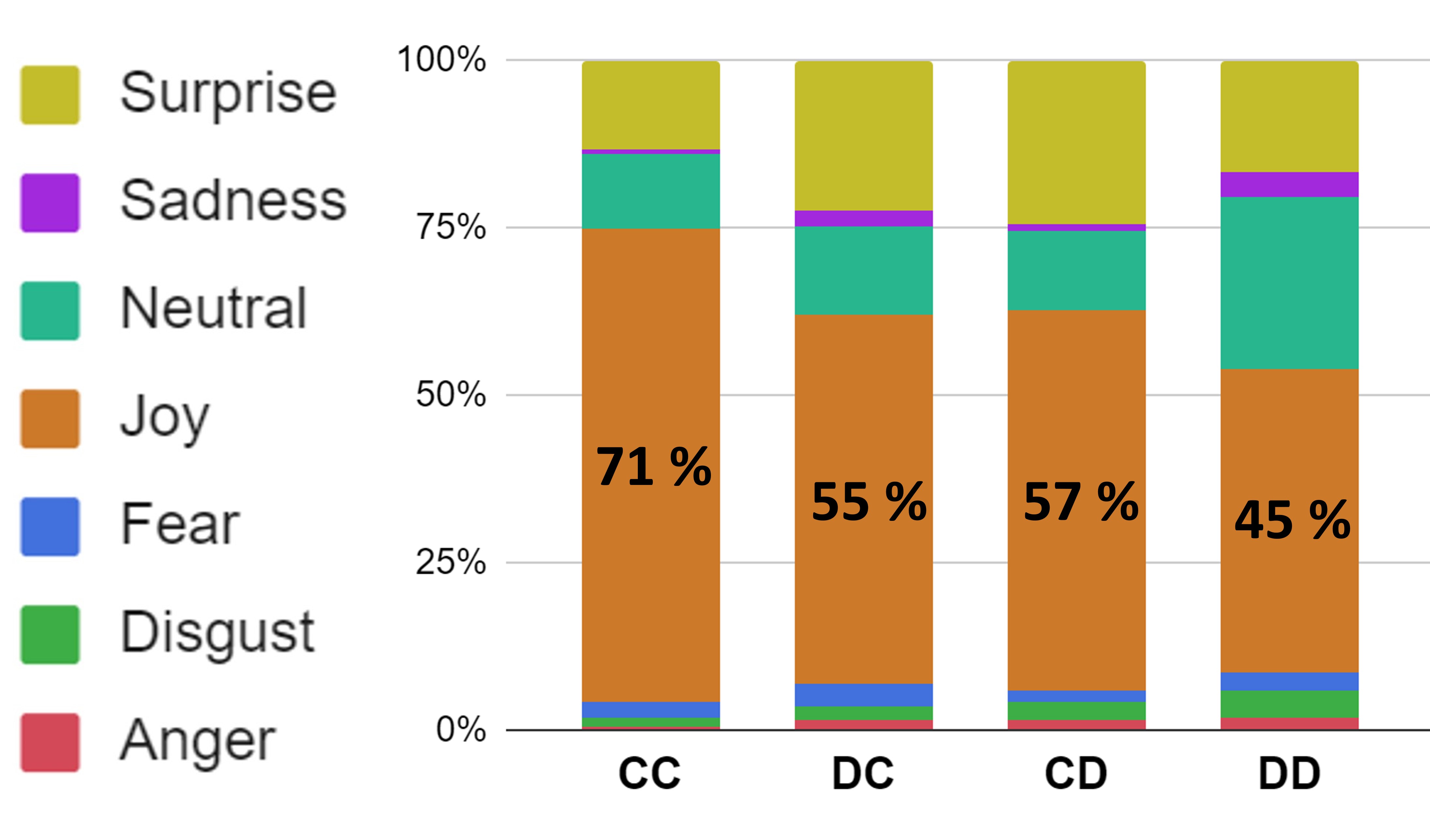}
        \caption{Context-free}
        \label{fig:zerocontext}
    \end{subfigure}
    \begin{subfigure}[b]{0.295\textwidth}
        \centering
        \includegraphics[width=\textwidth]{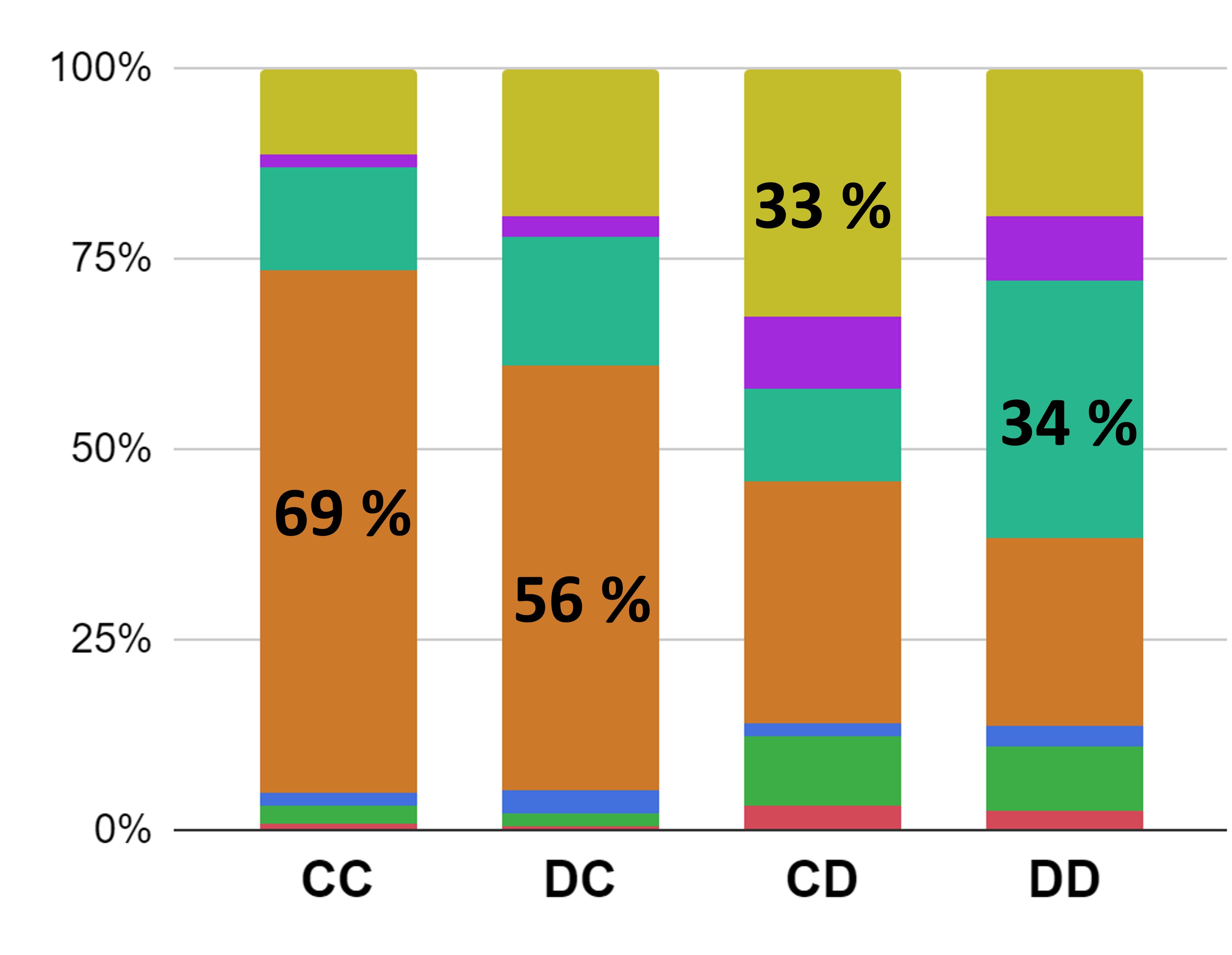}
        \caption{Context-based}
        \label{fig:withcont}
    \end{subfigure}
    \begin{subfigure}[b]{0.295\textwidth}
        \centering
        \includegraphics[width=\textwidth]{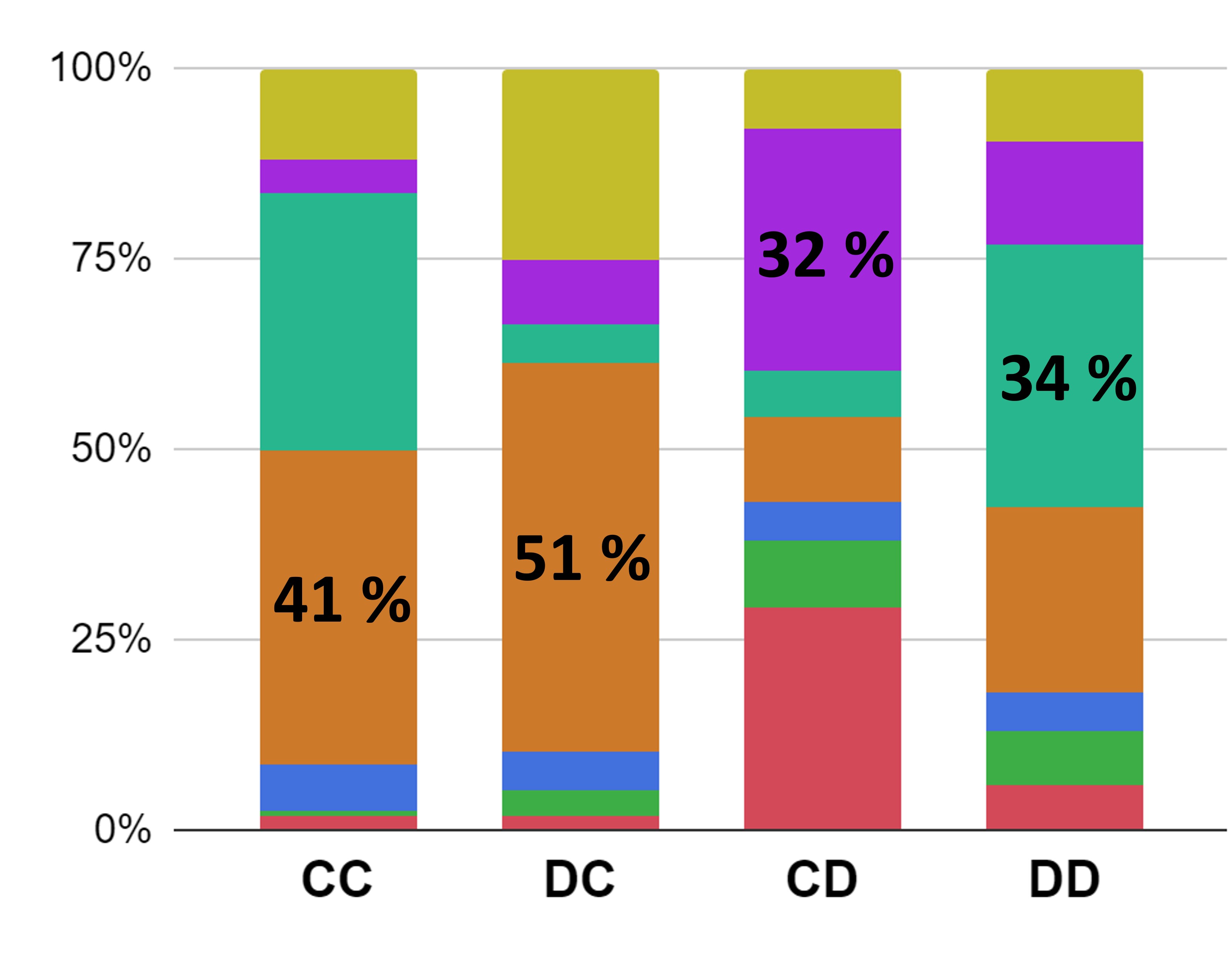}
        \caption{Context-only}
        \label{fig:human_novideo}
    \end{subfigure}
    \caption{(a) \textit{Context-free}, (b) \textit{Context-based}, (c) \textit{Context-only}. CC (mutual cooperation), DC (Player A exploit), CD (Player A exploited), DD (mutual defection).}
    \label{fig:three_graphs}
\end{figure*}

BCI predicts observers' beliefs about what someone feels, not the actual feelings of the person producing the expression. 
Thus, it is most applicable to predicting how the observer will behave in social settings~\cite{weisman2009being,van2007expressing}, though prior psychological research also suggests that context-based perceptions are more consistent with self-reported feelings than context-free perceptions~\cite{hess2022infusing}. 
As BCI predictions are expressed as a probability distribution over labels, rather than a specific class, BCI is aligned with recent innovations in affective computing that leverage annotator variability as crucial information for improving recognition accuracy~\cite{lotfian2017formulating,prabhu2023end}.

\subsection{Archival Data}

We first replicate the utility of BCI on a novel dataset, USC's Split-Steal corpus~\cite{rapoport1965prisoner, lei2023emotional}, before integrating it with automated methods. 
This is a large collection of participants that engaged in a 10-round  prisoner’s dilemma task. Participants could see each other but not speak and were incentivized by playing for lottery tickets for several \$100 USD lotteries. The corpus consists of 7-second “reaction shots” when players learn of their joint choice on a given round. 
In each round, players can choose to cooperate (C) or defect (D), where cooperating is an attempt to split ten lottery tickets and defect is an attempt to steal all the tickets. The possible outcomes depend on the joint choice: both choose to equally split the tickets (CC), one player successfully steals from their partner (DC), one player is stolen from (CD); or both attempt to steal from each other (DD) with each receiving a single ticket. 

For our replication, we chose 25 of the most expressive videos for each possible outcome in the game as we wanted to focus on how these expressions would alter inferences. The USC Split-Steal corpus includes automatically derived expressivity ratings and we simply selected the 25 most expressive clips for each possible outcome in the game. 
This resulted in a database of 100 7-second video clips, 25 from each of the four actual game outcomes: CC, DC, CD, and DD.

\subsection{Emotion Perception Ratings}
\label{subsec:perception-rating}

We augmented this corpus following the procedure of Ong and colleagues~\cite{ong2015affective}. We recruited multiple annotators to estimate probability distributions that correspond to the probability that an emotion is perceived from the face alone (context-free), from the context alone (context-only) and from the face and context together (context-based). Annotators were recruited through Amazon Mechanical Turk and pre-tested to filter inattentive annotators.  
For each video, annotators were asked to identify one of 6 basic emotion (or neutral) they perceived the person to be feeling. Basic emotions were chosen to allow direct comparison to prior BCI results and because these are a natural language for rating perceived emotion (though the approach is easily extended to other schemes).

Separate groups of annotators were recruited to judge the videos with or without context. Within each group, each annotator rated 10 randomly selected videos and twenty ratings were obtained for each video in each context. 

\textbf{Context-free}: 
For $P(e|f)$, annotators were asked to rate emotions only seeing the video and without being told any other information. The only thing that could be inferred from the background was they are sitting in a room full of computers (see Fig.~\ref{fig:context_level}). 
They were simply told to watch the video and answer the questions. 
They were free to watch the video as many times as possible.

\textbf{Context-based}: 
For $P(e|c,f)$, annotators first were provided a description of the game context, including the payoff for different choices, and quizzed to verify their understanding of the structure of the game. They then rated 10 videos, each showing the joint outcome of a round and the nonverbal reactions of the two players involved (see Fig.~\ref{fig:context_level})
For each video, annotators were instructed to evaluate the emotional reaction of the player (Player A) highlighted in the red box. They were allowed to watch the video multiple times to ensure an accurate assessment of the player's emotional expression.

\textbf{Context-only}: 
For $P(e|c)$, annotators first received the same description of the game and quiz provided in the context-based annotation task. Rather than seeing a video, they only saw a text description of one of the four possible game outcomes and were asked to predict what a player experiencing this outcome was likely to feel.  
We recruited 141 annotators from Amazon Mechanical Turk, of which 20 annotations were discarded due to failing the attention check. 

\subsection{Emotion Probability Distribution}

BCI adopts a probability distribution approach for emotion recognition, utilizing ``soft" labels to represent the probability of each emotion, rather than producing a single label (e.g., Anger) from a video. This method addresses the challenges of subjective perception in image emotion recognition, as highlighted in previous research~\cite{zhao2017approximating,joshi2011aesthetics}.
Significantly,  this approach is supported by research in emotion recognition. For example, Lotfian and Busso~\cite{lotfian2017formulating} formulated emotion perception as a probabilistic model, and Prabhu et al.~\cite{prabhu2023end} advocated for label uncertainty modeling in speech emotion recognition.
These studies highlight the importance of using a probabilistic perspective in affective computing to better capture the complexities of emotional expression.

\subsection{Aggregate Analysis of Emotion Ratings}

Fig.~\ref{fig:three_graphs} visualizes the elicited emotion probability distributions. 
For the sake of simplicity, rather than showing the human ratings for each individual video, we average across all the videos with the same game outcome.

For \textit{context-free} perceptions (Fig.~\ref{fig:three_graphs}a), Joy is the predominant emotion across all game outcomes, with the CC very likely to be rated as joyful (71\%), DD being perceived as the least joyful (45\%), and 
DC and CD falling in the middle. Despite the prevalence of joy, most individual videos failed to reach strong annotator consensus.
Table~\ref{tab:majority_class} shows the fraction of videos with a clear majority label (over 50\% agree on the label) and clear consensus (over 66\% agree on the label). 

Fig.~\ref{fig:three_graphs}b illustrates the context-based results, where annotators had knowledge of both facial cues and the game outcome. Joy was still prevalent in the CC condition at 69\%, and DC it was at 56\%. 
However, the CD condition was marked by a 33\% prevalence of Surprise, and DD by 34\% Neutral. These results indicate that emotions in the CD and DD conditions are not overwhelmingly dominated by a single emotion label but rather display a mixed (for instance, CD has 33\% Surprise and 32\% Joy), supporting the necessity of an emotion distribution approach for more precise emotion recognition.
Table~\ref{tab:majority_class} illustrates a diminished consensus for outcomes such as CD and DD, underscoring the insufficiency of a single-label approach. 
These results support that soft labeling strategies to more accurately capture the varied emotional nuances revealed when contextual information is incorporated.

\begin{table}[ht]
\centering
\large 
\resizebox{0.5\textwidth}{!}{
\begin{tabular}{lccc}

\thickhline
              & \textbf{Outcome} & \textbf{\%Majority class} & \makecell{\textbf{\%Supermajority} \\ \textbf{class ($\geqq$ 2/3)}} \\ 
\thickhline
Context-free & CC                & 0.92                      & 0.64                                \\
             & DC                & 0.72                      & 0.44                                \\
             & CD                & 0.80                      & 0.52                                \\
             & DD                & 0.72                      & 0.36                                \\
\midrule
Context-based & CC               & 0.92                      & 0.56                                \\
              & DC               & 0.72                      & 0.48                                \\
              & CD               & 0.24                      & 0.08                                \\
              & DD               & 0.44                      & 0.08                                \\
\bottomrule
\end{tabular}
}
\caption{Comparison of majority and super-majority class consensus in \textit{context-free} and context-based.}
\label{tab:majority_class}
\end{table}


In the context-only annotations (Fig.~\ref{fig:three_graphs}c), there was a notable overestimation of Sadness and Anger in the CD condition when compared to context-based perceptions. 
Again, multiple emotions were again prominent, as in CD, 32\% was Sadness and 29\% was Anger, which also supports the need for an emotion distribution approach. A more detailed evaluation will be discussed in Section 4.



\section{Automatic Approaches for Facial and Contextual Emotion Recognition}
As illustrated in Fig.~\ref{fig:overview}, we automate context-based predictions by (1) estimating $P(e|f)$ from facial videos, (2) estimating  $P(e|c)$ from a textual description of the game context, and (3) integrating these estimates. To demonstrate generality, we evaluate several alternative techniques at each stage. Additional details found in supplemental materials~\cite{han2024knowledge2}.

\subsection{$P(e|f)$ - Emotion Probability given by Face}

We compare three alternatives for automatically recognizing emotions from decontextualized videos. 
This allows us to compare the accuracy of different approaches but also to examine if knowledge-based recognition can benefit a range of methods. We evaluate a commercial approach (FACET) and a state-of-the-art pre-trained model (EAC model). 
Each of these methods recognizes emotions frame-by-frame and may miss important information encoded in the dynamics of the video. Thus, we also train a dynamic LSTM model that can make predictions based on changes in expressions within a video. In evaluating each method, we use treat \textit{Context-free} human annotations as ground truth (i.e., how well can each method predict what emotions are perceived by observers without access to the game context?).  

\textbf{1) FACET:} FACET is a commercial expression recognition method based on the Computer Expression Recognition Toolbox~\cite{littlewort2011computer} and produces a distribution of emotion labels for each frame of video. 
Specifically, FACET provides 'evidence values' for each frame, indicating the likelihood of an expression corresponding to a specific emotion, with values ranging from -4 to +4. We set negative values to zero, average the evidence across all video frames and re-scale the result to ensure that the sum of probabilities equals 1. 

\textbf{2) EAC:} Erasing Attention Consistency (EAC) model~\cite{zhang2022learn}, is a state-of-the art emotion recognition method based on  ResNet. EAC analyzes videos frame-by-frame and we extract and average emotion probabilities from the softmax layer to determine the overall emotional distribution for each video.
Specifically, we use a pre-trained EAC model with the RAF-DB dataset (Real-world Affective Faces)~\cite{li2017reliable}. The extracted probabilities are re-scaled to ensure that their sum equals 1.

\textbf{3) LSTM:} The previous two methods ignore how expressions change across the 7-second video, yet some emotion impressions, such as surprise, might arise from quick facial movements. To capture these, we train an LSTM model that incorporates dynamics features. The model is trained using human context-free annotations as the ground truth and utilizes a range of input features, including sequences of Action Units (AUs), facial optical flows, gaze, and head pose data.

We employ OpenFace 2.0~\cite{baltrusaitis2018openface} to extract 12 Facial Action Units\footnote{The selected AUs are AU 1, 2, 4, 6, 7, 10, 12, 14, 15, 17, 25, and 26.}, focusing on those commonly co-occurring AUs identified by prior work~\cite{stratou2017refactoring}. 
For optical flows, the ZFace tool~\cite{jeni2015dense} is used to track the movement of dense facial landmarks (512 points) over time, allowing us to calculate the flow between each frame. 
This data helps the LSTM model to capture the dynamic expressions and subtle changes in the face that are crucial for dynamic emotions. 
Additionally, OpenFace provides gaze direction vectors and gaze angles.
Also, head pose direction vectors are provided.
To train the model, we pre-process and regularize the input.
These include the normalization of input features to reduce potential bias, the incorporation of dropout layers to prevent overfitting, and scaling to accelerate the convergence of the training process. 
The validation strategy employed is Leave-One-Out Cross-Validation (LOOCV), providing a thorough assessment of the model's performance. Note, that as LSTM is fine tuned on the Split-Steal corpus, whereas the other approaches are pre-trained, LSTM can be seen also as an attempt to create an upper-bound on the accuracy of context-free judgments.


\subsection{$P(e|c)$ - Emotion Probability given by Context}

We compare GPT-3.5, GPT-4~\footnote{GPT-3.5 and GPT-4 versions as of February 1, 2024}, Llama 2, and Gemini~\footnote{Llama 2 and Gemini versions as of May 15, 2024} LLM models for their ability to infer likely emotions from a textual description of situational context.
Models are given the identical descriptions and questions that were provided to the human annotators, with minor changes to get the models to produce standardized outputs  (see Fig~\ref{fig:gpt_prompt}). 
The prompt contains three main components:
\begin{itemize}
    \item \textcolor{Color1}{A general description of the prisoner's dilemma game.}
    \item \textcolor{Color2}{The game outcome of each turn (CC,DC,CD, and DD).}
    \item \textcolor{Color3}{A request for the emotional distribution (Basic emotion).}
\end{itemize}

\begin{figure}[h]
\begin{tcolorbox}[colback=white,colframe=black,title= GPT Prompt - $P (e|c)$]
\textcolor{Color1}{
Imagine a scenario where two people, Player A and Player B, play a competitive game called ``Split or Steal.”
Players play multiple rounds with each other. In each round of the game, players each decide whether to split or steal from a pot of \$10. If both choose ``split”, they each get \$5.
``If both choose ``steal”, they each get \$1. If one chooses ``split” but the other chooses ``steal”, the stealer gets all \$10. They make their choices secretly and their choices are revealed at the end of the round. 
Scenarios describe one round of the game. Imagine the feelings of Player A.}
\textcolor{Color2}{
In this round, Player A chooses ``steal" and Player B chooses ``split."}
\textcolor{Color3}{
How does Player A experience emotions? Provide a probability distribution based on the following emotion list: Joy, Neutral, Surprise, Anger, Disgust, Fear, Sad. Ensure that the sum of probabilities is 1.
Provide answer in the following format:
``Joy: {prob 1}, Neutral: {prob 2}, Surprise: {prob 3}, Anger: {prob 4}, Disgust: {prob 5}, Fear: {prob 6}, Sad: {prob 7}.”}
\end{tcolorbox}
\caption{GPT Prompt depicting a ``Split or Steal" game scenario used to $P(e|c)$ data on emotion probability distribution.}
\label{fig:gpt_prompt}
\end{figure}

For evaluation, human context-only annotations are used as ground truth (mentioned in Section~\ref{fig:human_novideo}). 
We set the temperature parameter of each LLM model to the default when conducting the experiment. 
Each version was prompted 20 times for each description, and the results were averaged. 
\subsection{Best method for $P(e|f)$ and $P(e|c)$}

\begin{table}[b]
    \centering
    
    \setlength{\tabcolsep}{3.5pt} 
    \renewcommand{\arraystretch}{1.1} 
    \resizebox{0.49\textwidth}{!}{%
    \begin{tabular}{
    >{\centering\arraybackslash}m{1.0cm}
    >{\centering\arraybackslash}m{1.1cm}||
    >{\centering\arraybackslash}m{1.1cm}
    >{\centering\arraybackslash}m{1.1cm}
    >{\centering\arraybackslash}m{1.4cm}
    }
    
    \thickhline
    \textbf{Face} & \textbf{Context} & \textbf{RMSE(↓)} & \textbf{KLD(↓)} & \textbf{F1-score(↑)} \\
    \hline
    FACET & - & 0.119 & 0.688 & \textbf{0.689}\\

    EAC & - & 0.155 & 2.421 & 0.658 \\
    LSTM& - & \textbf{0.095} & \textbf{0.134}  & 0.660 \\
    \hdashline
    - & GPT-3.5 & 0.197 & 2.579&0.333  \\
    - & GPT-4 & \textbf{0.061} & \textbf{0.101}& \textbf{0.750} \\
    - & Llama 2 & 0.149 & 0.343 & 0.1  \\
    - & Gemini & 0.137 & 0.491 & 0.333  \\
    \hline
    \end{tabular}
    }
    \caption{$P(e|f)$ and $P(e|c)$, showing the match between automated predictions and the distribution of human \textit{context-free} and \textit{context-only} perceptions (lower RMSE, KLD indicate a better match, while a higher F1 indicates better accuracy)}
    \label{tab:Result_face_context}
\end{table}

We next compare the performance of different models for de-contextualized facial emotion recognition ($P(e|f)$) and context-based emotion recognition ($P(e|c)$). 

\textbf{Evaluation Metrics:} 
To assess the performance of our models, we employ three standard metrics. Kullback-Leibler divergence (KLD)~\cite{kullback1951information} and Root Mean Square Error (RMSE)
are standard metrics to compare the distance between two probability distributions~\cite{peng2015mixed,zhao2017approximating}, 
which is most appropriate given the variability in labels provided by the human annotators.  
KLD directly measures the discrepancy between two probability distributions, with lower values indicating better model performance. 
RMSE provides a measure of the average magnitude of the errors, again with lower values being preferable. Finally, we include F1 (weighted) to assess performance if the model was forced to provide a single label, though caution that F1 can be misleading as only about half of the videos had strong agreement amongst the annotators.

\textbf{Best model for $\boldsymbol{P(e|f)}$:} The LSTM model is distinguished by the lowest RMSE (0.095) and KLD (0.134), suggesting strong predictive accuracy and a high correlation with human-annotated data for capturing facial emotion nuances (as referenced in Table~\ref{tab:Result_face_context}). While the LSTM model is fine-tuned for IPD dataset, our emphasis is not on its distinct advantages. 
Rather, we focus on demonstrating how BCI enhances a range of facial emotion recognition methods.

\textbf{Best model for $\boldsymbol{P(e|c)}$:} GPT-4, with the lowest RMSE (0.061) and KLD (0.101), indicates that its predictions for context-based emotions closely mirror human judgments (as seen in Table~\ref{tab:Result_face_context}).
These results support our decision to employ LSTM for facial emotion recognition and GPT-4 for contextual emotion recognition for integration model $P(e|c,f)$, which promises to align closely with human emotional perception.

\section{Integration of Facial and Contextual Emotion Recognition}

We compare alternative methods for integrating facial and contextual cues for context-aware emotion recognition.
We explore two methods: BCI and GPT-4 Integration method.

\textbf{1) Bayesian Cue Integration (BCI):}
We apply BCI as detailed in Eq.~\ref{eq:bayesian}, using results of $P(e|f)$ from three facial emotion recognition methods~\footnote{FACET, EAC, and LSTM} and $P(e|c)$ from GPT-3.5 and GPT-4. Following \cite{ong2015affective}, we do not explicitly calculate $P(e)$ but normalize the product into a proper distribution by dividing each probability by the sum of the posteriors.

\textbf{2) GPT-4 Integration:}
We explore if GPT-4 by itself could  integrate facial cues and contextual information to generate $P(e|c,f)$. 
This method leverages the advanced capabilities of GPT-4 to directly generate a context-aware emotion probability distribution. 
The integration process involves GPT-4 with a representation of $P(e|f)$ estimated by a context-free facial emotion recognition method. 
Based on previous research~\cite{sui2023evaluating,wang2023exploring}, LLMs have a better understanding when the input is represented in natural language rather than numerical value. 
Therefore, we add natural language descriptions corresponding to the probability levels to enhance the model's interpretability. 
For example, a Joy probability above 0.5 would be described as \textit{``a high level of happiness"} within the input prompt. 


\subsection{Overall Result}

We compare the performance of knowledge-based recognition with the distribution of human context-based perceptions (see Table~\ref{tab:comparison_results}).  Two broad observations are immediately clear.  First, GPT-4 generally yielded better performance when compared with GPT-3, though this benefit was strongest in combination with LSTM. Second, LSTM (which incorporated dynamic facial movements), clearly dominated the other context-free methods in predicting the distribution of human perceptions (even exceeding the performance of BCI using human labels as measured by KLD).

Looking in more detail, using GPT-4 to predict $P(e|c)$ yielded strong improvements in performance with LSTM across all three measures of performance, when compared with GPT-3. For FACET and EAC, performance was improved in only two of the three measures. This is interesting as GPT-4 was far better at capturing human context-only perceptions, suggesting there is some interaction between errors in the context-based distribution when combined with the context-free judgments in these two methods. An analysis of the confusion matrices of EAC and FACET suggests they were harmed by their inability to recognize surprise. 

GPT-4 with BCI (denoted as LSTM+GPT-4 (BCI)) showed essentially equivalent performance with BCI using human perceptions as measured by KLD and RMSE, but not for F1. 
Note that KLD and RMSE capture the closeness between two distributions. 
In contrast, F1 forces the model to pick the most likely class, even if another class was almost equally likely. 
To understand the difference, we examine the individual videos and found that the difference in accuracy between LSTM+GPT-4 (BCI) and BCI with human distributions is due to differences in videos where the player was exploited (CD). 
Most of these videos showed an almost equivalent likelihood of being labeled as joy or surprise but the human context-based labels tended to assign somewhat weight to joy, whereas LSTM+GPT-4 (BCI) assigned more way to surprise. This highlights the problematic consequences of using hard labels when the probability of perceiving multiple classes is high.

Interestingly, using GPT-4 to perform the integration, in addition to reasoning about the situational context, yielded remarkably strong results. This approach actually improved over BCI for FACET and EAC, but yielded slightly worse performance than BCI with LSTM. A disadvantage of GPT-4 integration is this is a black box so it more difficult to gain insight into why the method produced improvements or deficits. Nonetheless, this suggests that there is promise it using LLMs as a replacement for the BCI approach.

We also look at improving integration with a nonlinear approach (training a Neural Network), but this performed somewhat worse than BCI and LSTM (GPT-4) (see supplemental materials~\cite{han2024knowledge2}).

\begin{table}[h]
    \centering
     \setlength{\tabcolsep}{3.5pt}  
     \renewcommand{\arraystretch}{1.05} 
    \resizebox{0.490\textwidth}{!}{%
    \begin{tabular}{l||ccc}
     \thickhline
    \textbf{Face+Context (Integration)} & \textbf{KLD(↓)} & \textbf{RMSE(↓)} & \textbf{F1(↑)} \\
    \hline
    FACET+GPT-3 (BCI) & 1.713 & 0.215 & 0.525 \\
    FACET+GPT-4 (BCI) & 1.829 & 0.200& 0.565 \\
    EAC+GPT-3 (BCI) & 1.340 & 0.200 & 0.519\\
    EAC+GPT-4 (BCI) & 1.330 & 0.210 &0.527\\
    LSTM+GPT-3 (BCI) & 0.809 & 0.162& 0.454\\
    LSTM+GPT-4 (BCI) & \textbf{0.346} & 0.104&0.649 \\
    Human+Human (BCI) & 0.441 & \textbf{0.092}& \textbf{0.782} \\
    \hdashline
    FACET (GPT-4) & 0.648  & 0.150&0.528 \\
    EAC (GPT-4) & 0.597 & 0.155&0.503 \\
    LSTM (GPT-4) & 0.354 & 0.112&0.530 \\
    \hdashline
    LSTM+GPT-4 (NN) &  0.580  &  0.151 &  0.151 \\
    \hline
    \end{tabular}
    }
    \caption{Comparison of alternative integration methods for $P(e|c,f)$ using different combinations of facial (FACET, EAC, LSTM) and context (GPT-3 and GPT-4) and integration methods (BCI and GPT-4). The best results for each measure.}
    \label{tab:comparison_results}
\end{table}

\vspace{-0.7em}

\subsection{How Integration improves Performance?}

\begin{figure}[h]
    \centering
    \includegraphics[width=0.87\linewidth]{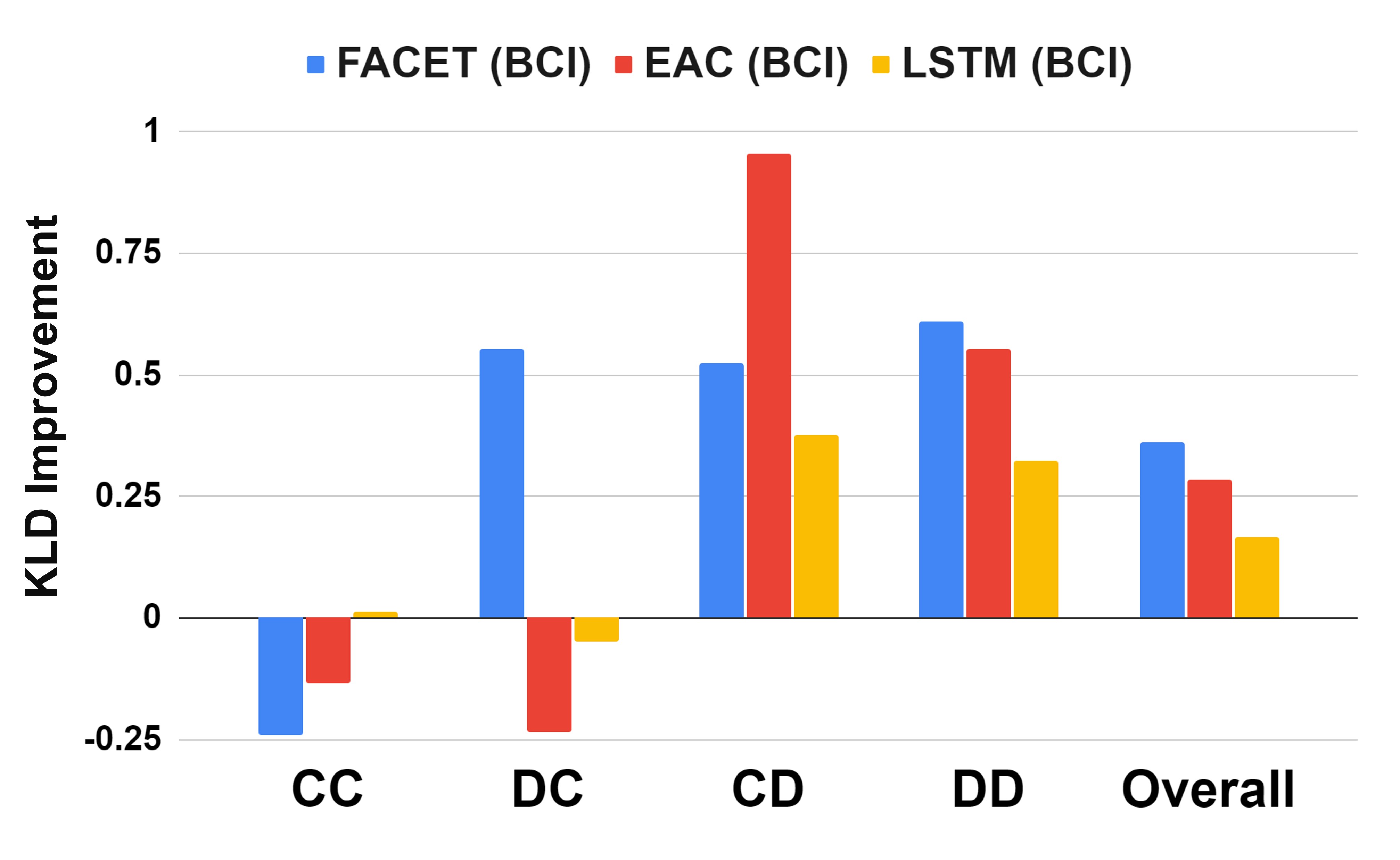}
    \caption{Enhancement in model performance (KLD) through facial and context integration.}
    \label{fig:kld-context}
\end{figure}

We perform further analysis to examine how BCI improves recognition performance.
Fig.~\ref{fig:kld-context} illustrates the change in recognition performance as a function of the different game outcomes. Specifically, it shows the change distance between context-free and context-based estimates that result from using BCI with GPT-4 (measured by the change in KLD). 
 Positive numbers indicate improved performance. This figure indicates that all methods improved their performance in predicting perceived emotions when the game outcome was disadvantageous to the player. For example, whereas a context-free method might predict joy, learning the person was exploited might change this to surprise). This did come at some cost in that methods became somewhat worse at predicting emotions when the game outcome was advantageous to the player. Across all game outcomes, performance improved.
This emphasizes the significant role that context plays in accurately interpreting emotional perception across various integration methods.

\vspace{-0.5em}
\section{Conclusion}

In this study, we explored the potential benefits of knowledge-based emotion recognition. Inspired by a psychological theory of human emotion perception -- Bayesian Cue Integration -- we find that context-free automatic recognition improved (across all methods tested) by incorporating zero-shot inferences from LLMs about the situational context. Specifically, we evaluated methods using naturalistic expressions produced during a two-person prisoner's dilemma task and found that the integration of facial cues and contextual information using BCI accurately predicted the ratings by human annotators knowledgeable of the situational context. Performance improved across all methods, with the LSTM and GPT-4 combination achieving the best performance (though it should be noted that LSTM was trained on the Split-Steal corpus while other methods used pre-trained models).

Knowledge-based recognition showed the strongest improvements when the player experienced a negative outcome (i.e., they were exploited by their partner or both partners tried to exploit each other). This seems to be because players often showed smiles that, in the absence of context, were interpreted as joy but when seen in the light of context were interpreted more negatively. For example, in one anecdote, a player can be seen mouthing profanity at her partner while wryly smiling (a detail missed by annotators without access to the context). In contrast, context failed to improve the accuracy of emotion predictions when the player experienced a positive outcome. This seems to be because players almost always showed some smile as a result of the outcome. Together, these emphasize the accuracy of context-free emotion recognition will depend heavily on the context, highlighting that the integration of situational context as a vital component in interpreting emotional states, particularly in complex social situations. Thus, findings highlight the promise of knowledge-based approaches as a direction for future research in the field of affective computing.

Our findings further reinforce prior studies that highlight the zero-shot social and emotional intelligence of LLMs. Without any fine-tuning, GPT-4 showed consistency with the emotions reported by humans given a description of an emotional situation. While here we only tested on a single task, in light of other studies, this suggests a robustness and adaptability to various domains and datasets. This universality is an advantage, broadening the applications of our approach across different areas of research and practical deployment.

The findings of this study open several avenues for future research in emotion recognition and affective computing. While our results show the potential of integrating facial expressions and situational context using BCI and LLMs, there is still room for refinement in closing the gap between automated methods and human emotion perception.
BCI uses a simplified model of human emotion perception, which might not fully account for situations where display norms constrain emotional expression. For example, research on the prisoner's dilemma~\cite{hoegen2023expression} reveals discrepancies between first-person and second-person reports of emotion, likely due to emotion regulation strategies not captured by BCI. Additionally, our focus was on perceived emotion, but we did not examine how well these perceptions align with actual feelings.

Culture may also play a role in our findings. Research suggests that the influence of others' facial expressions is strongest in interdependent cultures~\cite{hess2016judging}. Given that our annotators were US-based and the culture of the US is low in interdependence, this might explain our results. 
Moreover, LLMs have been noted for their anglocentric tendencies, which may impact their efficacy in interpreting emotions from diverse cultural contexts~\cite{havaldar2023multilingual}. 
This highlights the need for caution in extrapolating our findings to other tasks, contexts, or cultures.

Moreover, the use of probabilistic programming could offer a more sophisticated approach to modeling the complex interplay between facial expressions, context, and emotion perception~\cite{ong2019applying}. 
By developing more advanced probabilistic models, researchers could further reduce the gap between automated emotion recognition and human-like emotion understanding.

Our method shows promise for structured scenarios with clear outcomes, like poker games or negotiations, where outcomes can be easily described for LLM prompts. However, extending this method to more unfolding scenarios, like depression interviews, poses challenges due to the difficulty in describing events in video in real-time for LLM prompts.

Finally, though BCI has shown promise in explaining human context-based predictions across a range of settings, our current study only explored a single corpus. 
While it is encouraging that our computational findings mirror (and replicate) these psychological findings, future research must validate and extend our findings to other social tasks.

\section*{Ethical Statement}

This paper examined how people form judgments of emotional expressions through re-analysis of previously collected data subject to ethical review and shared based on consent terms. The data was annotated with human coders and analyzed using commercial automatic facial analysis methods. These decisions introduce bias and limit result generality. The original data is demographically diverse but collected only in Los Angeles. Perceived-emotion judgments were obtained via Amazon Mechanical Turk workers. Automated methods have biases in tracking and characterizing people of color. Data was limited to the prisoner’s dilemma game, necessitating replication across tasks, populations, and facial analysis tools for robust conclusions.

\section*{Acknowledgment}
This work is supported by the Army Research Office under Cooperative Agreement Number W911NF-20-2-0053. The views and conclusions contained in this document are those of the authors and should not be interpreted as representing the official policies, either expressed or implied, of the Army Research Office or the U.S. Government. The U.S. Government is authorized to reproduce and distribute reprints for Government purposes notwithstanding any copyright notation herein.

\bibliography{reference}
\bibliographystyle{IEEEtran}

\end{document}